\documentclass[aps,pra,twocolumn,showpacs,superscriptaddress,preprintnumbers,amsmath,amssymb]{revtex4}
\usepackage{graphicx}
\usepackage{dcolumn}
\usepackage{bm}
\usepackage{amsmath}

\newcommand{\bra}[1]{\langle #1|}
\newcommand{\ket}[1]{|#1\rangle}

\begin{document}
\title{Quantum logic via the exchange blockade in ultracold collisions}
\author{David Hayes}
\email{dh123@unm.edu}
\affiliation{{Department of Physics and Astronomy, University of New Mexico, Albuquerque, New Mexico 87131, USA}}
\author{Paul S. Julienne}
\affiliation{{Atomic Physics Divsion, National Institute of Standards and Technology, Gaithersburg, Maryland 20899-8423, USA Atomic Physics Divsion}}
\author{Ivan H. Deutsch}
\affiliation{{Department of Physics and Astronomy, University of New Mexico, Albuquerque, New Mexico 87131, USA}}
%
\begin{abstract}
A nuclear spin can act as a quantum switch that turns on or off ultracold collisions between atoms even when there is neither interaction between nuclear spins nor between the nuclear and electron spins.  This ``exchange blockade" is a new mechanism for implementing quantum logic gates that arises from the symmetry of composite identical particles, rather than direct coupling between qubits.  We study the implementation of the entangling $\sqrt{\text{SWAP}}$ gate based on this mechanism for a model system of two atoms, each with ground electronic configuration $^1S_0$, spin 1/2 nuclei, and trapped in optical tweezers.  We evaluate a proof-of-principle protocol based on adiabatic evolution of a one-dimensional double Gaussian well, calculating fidelities of operation as a function of interaction strength, gate time, and temperature.  
\end{abstract}
\maketitle
Implementations of quantum processors require coherent quantum logic gates that induce entangling operations between qubits~\cite{Nielsen}.   In most cases, the same physical effect that gives rise to coherent couplings in the logical basis states leads to decoherence between them.  For example, in an ion trap, the strong Coulomb interaction provides the quantum bus for two-qubit logic \cite{Cirac}, but charged ions can be heated by fluctuating patch potentials in trap electrodes \cite{Wineland}.  In the neutral atom platform, strong spin dependent collisions can lead to an entangling gate \cite{Jaksch1, Bloch, Stock} but also to spin relaxation \cite{Stoof}. 

In this article, we propose a new scheme for quantum logic based on an ``exchange blockade" arising solely from the symmetry of identical composite particles rather than from differential coupling strengths.  We consider a hybrid approach based on NMR and ultracold collisions of trapped neutral atoms.  Elements with two valence electrons in a closed shell configuration $^1S_0$, such as those in group-II, possess no electron spin and thus no hyperfine interaction with nuclear spin when in the electronic ground state.  Nuclear spin, nonetheless, strongly affects the interaction between these atoms solely due to quantum statisitics.  In a fermionic species, the dominant ultracold s-wave collisions are forbidden by the Pauli exclusion principle when the nuclei are in a symmetric spin state, and are allowed only for the nuclear-spin antisymmetric states.  The single-channel cold collisions, governed by {\em electronic interactions} between atoms, thus induce an exchange interaction between {\em nuclei}, even though there is no direct interaction between nuclei nor between electrons and nuclei. The ability of nuclear spins to act as a switch that turns on and off electron interactions provides a mechanism by which quantum information can be protected from the environment while simultaneously qubits are strongly coupled to one another in a manner depending on the logical basis states.

We take group-II-like elements with isotopes of spin $I=1/2$ nuclei to encode qubits, e.g. $^{171}$Yb, with logical basis states defined in the usual way, $\ket{0}=\ket{\uparrow}$, $\ket{1}=\ket{\downarrow}$.  Single qubit rotations can be achieved via rf-pulses in a strong bias magnetic field, as in NMR.  Interactions between atoms that have nuclei spin polarized in the same direction are blockaded (to the degree that $p$-wave collisions are negligible).  Atoms undergo an exchange interaction as the nuclear-spin singlet component receives an elastic collisional phase shift relative to the noninteracting triplet component, resulting in the well known $\sqrt{\text{SWAP}}$, a universal entangling gate \cite{Nielsen}.  The exchange interaction was previously studied in the seminal work of Loss and DiVincenzo in the context of electrons in a double quantum dot \cite{Loss}.  In our realization, the composite nature of the atom allows one to separate the degree of freedom which stores the quantum information (the nucleus) from that which provides physical coupling (the electrons).  
	
Coherent collisions require well-localized atomic wave packets whose motion is highly controlled.  We evaluate the performance of two-qubit entangling gates using dipole traps formed by tightly focused optical tweezers, recently used to trap individual atoms \cite{Grangier}. Dorner {\em et al.} have studied ultra-cold collisions of bosonic alkalis in such tweezers \cite{Calarco}.  Similar gates have been studied theoretically in a lattice of double-well potentials formed by multiple frequencies \cite{Charron} and more recently, experiments explore the control of atoms in double-well lattices formed in 2D through overlapping lattices of orthogonal polarizations \cite{Porto}.  We revisit the problem of implementing gates in double-well potentials, here in the context of the exchange blockade.

We take a simple one-dimensional model assuming tight confinement in directions transverse to the direction of the collision.   We model the two-tweezer dipole trap as a symmetric double well of Gaussian shaped potentials, each of depth $V_0$ and rms width $\sigma$, separated by distance $d$.  Single atom orbitals are the eigenstates of the double well, which are then filled with two atoms for a two-qubit system, in analogy to the two-electrons occupying the molecular orbitals of a diatomic hydrogen molecule.
The two atomic qubits interact via a contact term arising solely from s-wave collisions, $V_{\text{int}}(x_1-x_2) = g \delta(x_1-x_2)$. The coupling constant, $g$, follows from a quasi-1D approximation to the scattering process, ignoring renormalization effects \cite{Olshanni}. Throughout, we choose harmonic oscillator units such that energy is measured in units of $\hbar \omega$, where $\omega = \sqrt{V_0 /(m \sigma^2)}$ is the oscillation frequency along $x$, and lengths are measured in units of $x_0 = \sqrt{\hbar/(m \omega)}$.  The dimensionless interaction strength is then $g = 2 (a_s / x_0)(\omega_{\perp}/\omega)$, where $a_s$ is the $s$-wave scattering length and $\omega_{\perp}$ is the oscillation frequency in the transverse dimension. We take a well depth of $V_0 =10 \hbar \omega$, or $\hbar \omega = 20 E_{R}/(k \sigma)^2$, where $ E_{R}$ is the recoil energy.  Though in principle deeper wells and larger oscillation frequencies are possible \cite{Grangier}, high fidelity operation requires optimal control of that depth and an asymmetric ``tilt" of the well \cite{Calarco, Porto}, not under consideration here.  

To implement a gate, we envision separated noninteracting atoms, each in its own tweezer, that are brought together into the same well as $d \rightarrow 0$ where cold collisions occur and then are separated again into individual traps.  As the wave packets overlap, atoms lose their identity, and proper Fermi symmetrization of the state is necessary.  We define our two-qubit logical basis such that the first qubit is localized in the left ($L$) well and the second in the right ($R$),
\begin{subequations}
\begin{align}
\ket{0,0}&= \hat{f}^{\dagger}_{L,\uparrow}\hat{f}^{\dagger}_{R,\uparrow}\ket{vac} = \Psi_-(x_1,x_2) \ket{\uparrow \uparrow}, \\
\ket{0,1}&=\hat{f}^{\dagger}_{L,\uparrow}\hat{f}^{\dagger}_{R,\downarrow}\ket{vac} \\  
&= \frac{1}{\sqrt{2}}  \left( \Psi_-(x_1,x_2) \ket{\chi_T}+ \Psi_+ (x_1,x_2) \ket{\chi_S} \right) ,\nonumber  \\
\ket{1,0}&=\hat{f}^{\dagger}_{L,\downarrow}\hat{f}^{\dagger}_{R, \uparrow}\ket{vac} \\  
&= \frac{1}{\sqrt{2}}  \left( \Psi_-(x_1,x_2) \ket{\chi_T}- \Psi_+ (x_1,x_2) \ket{\chi_S} \right) ,\nonumber  \\
\ket{1,1}&= \hat{f}^{\dagger}_{L, \downarrow}\hat{f}^{\dagger}_{R, \downarrow}\ket{vac} =  \Psi_-(x_1,x_2) \ket{\downarrow \downarrow}.  
\end{align}
\label{logicalbasis}
\end{subequations}
Here we have used both first and second quantized notation, where $ \hat{f}^{\dagger}$ are fermionic creation operators for localized atoms,  $\Psi_{\pm}(x_1,x_2)=\left(\psi_L(x_1)\psi_R(x_2) \pm \psi_R(x_1)\psi_L(x_2) \right)/\sqrt{2}$ are two-atom orbitals, and $\ket{\chi_{T(S)}}=\left(\ket{\uparrow}\ket{\downarrow} \pm \ket{\downarrow}\ket{\uparrow}\right)/\sqrt{2}$ are the triplet(singlet) nuclear spin states with zero projection of angular momentum. The exact motional state is irrelevant to the encoded quantum information in the nuclear spin, but may affect the nature of the collision and fidelity of the gate. We begin by taking the atoms to be cooled to their motional ground state and later treat the effects of finite temperature. 

As seen in the logical basis states, Eq. (\ref{logicalbasis}), a SWAP operation, $\ket{0,1} \Leftrightarrow \ket{1,0}$, follows from a sign change between singlet and triplet components, or a $\pi$ rotation in this subspace; the entangling $\sqrt{\text{SWAP}}$ is a $\pi/2$ rotation. As a first attempt to implement this gate, we consider adiabatic evolution, though certainly this is not optimal for speed.  In order to better understand this, consider the case of noninteracting atoms.  Single particle eigenstates are parity eigenstates, which for large separation, become doubly degenerate pairs of states corresponding to gerade and ungerade combinations of the left/right localized wave packets in the separate tweezers denoted $\ket{g_n(u_n)} = \left( \ket{L_n} \pm \ket{R_n}\right)/ \sqrt{2}$ for the $n^{th}$ vibrational eigenstates.  For zero separation, $\ket{g_n} \rightarrow \ket{v_{2n}}$, $\ket{u_n} \rightarrow \ket{v_{2n+1}}$ , where $\ket{v_n}$ is a vibration eigenstate of the unified well. At large separation and in the absence of interactions, there is a four-fold degenerate ground-state manifold of two-particle orbitals.  These are described by the spatial orbitals with one atom per well as defined in the logical basis $\Psi_{\pm}$, and those with two-atoms per lattice site (associated with spin singlets),  $\ket{\Psi_c}= (\ket{L} \ket{L}+\ket{R}\ket{R})\sqrt{2}$ and  $\ket{\Psi_d}= (\ket{L} \ket{L}-\ket{R}\ket{R})\sqrt{2}$. The state $\ket{\Psi_d}$ has odd parity under spatial inversion and must be antisymmetric under spin exchange, and thus does not couple to the even-parity spin singlet in the logical basis.
  
In the presence of interactions, degeneracy between singly and doubly occupied states is broken.  Localized states become eigenstates for separated traps as in the Mott insulator phase \cite{Jaksch1,Greiner}. The degeneracy of states $\ket{\Psi_+}$ and $\ket{\Psi_c}$ for separated wells is broken by the onsite interaction. A relative phase then develops between singlet and triplet components of the logical basis, and the desired $\sqrt{\text{SWAP}}$ operation is achieved when, 
\begin{equation}
\phi_+ -\phi_-= \pi (1/2+n),
\label{adiabaticcondition}
\end{equation}
where $\phi_{\pm} =\int_0^t dt' E_{\pm}(t')/\hbar$ is the adiabatic dynamical phase.
\begin{figure}[h] 
   \centering
   \includegraphics[width=3.5in]{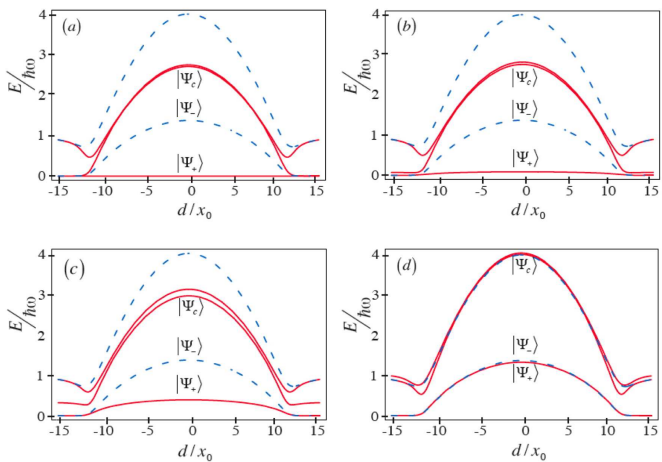} 
   \caption{Adiabatic energy curves relative to the noninteracting ground state as a function of trap separation for the first two triplet (blue dotted lines) and the first three singlet states (red solid lines) and for different values of interaction coupling strength $g$: (a) $g=0$, (b) $g=0.2$, (c) $g=1$, (d) $g=10$.  For large separation, these asymptote to single well energy levels with 0 or 1 unit of vibration.  All quantities are normalized to harmonic oscillator units (see text).  For large interaction strength, (d), singlets and triplets become degenerate corresponding to the Tonks-Giradeau regime.}
   \label{fig1}
\end{figure}

Adiabatic energy curves (relative to the noninteracting ground state singlet adiabatic curve) for states that have no more that one quantum of vibration in separated wells are shown in Fig. 1, for different interaction strengths.  Only even-parity singlets and odd-parity triplets are shown since these are the symmetries of the logical basis states.  The triplet states are unchanged by interactions and singlets are found by diagonalizing the s-wave contact interaction in the noninteracting basis.  The basis orbitals are calculated using a DVR method \cite{DVR} to find single-atom bound states in the double well. These are occupied with two atoms in accordance with Fermi symmetry.  

The behavior of the adiabatic curves as a function of interaction strength elucidates the possible operating points of a high-fidelity entangling interaction.  For weak interactions, $g \ll 1$, the very small energy gap between $\ket{\Psi_+}$ and $\ket{\Psi_c}$ limits the ability to maintain adiabaticity  as the traps are merged. In this regime, the tunneling energy dominates over the interaction energy before the wells are fully merged, and diabatic transitions arise as the ground state wave function changes from right/left localized to delocalized across the double well.   In principle, tunneling can be suppressed through an asymmetric well ``tilt" that provides stronger symmetry breaking, with the capacity to maintain adiabaticity while simultaneously inducing an exchange interaction \cite{Calarco,Porto}.  For intermediate interactions, $g \sim 1$, a large gap is opened between  $\ket{\Psi_+}$ and $\ket{\Psi_c}$ , in which case symmetric double wells is the preferred geometry for adiabatic operation of the gate.  For $g \gg 1$, whereas one might expect the most favorable operation of the exchange blockade, paradoxically the opposite is true. Singlet and triplet potentials again become degenerate as one enters the Tonks-Girardaeu regime of the one dimensional potential \cite{Tonks}.  In a contact interaction with infinitely strong coupling in one dimension, the particles are excluded from being at the same position, regardless of the symmetry of their spin state.  As a result, exchange interaction between spins is reduced.  This is illustrated in Fig. 2 which shows the gate time in the case of adiabatic evolution as a function of interaction strength for different values of $n$ in Eq. (\ref{adiabaticcondition}).  The divergence of $\tau$ as $g \rightarrow \infty$ represents the Tonks regime.  Since the Tonks gas is an artifact of our one-dimensional model, this is not a fundamental limitation.  In future work we will study tight confinement in 3D. 
\begin{figure}[htbp] 
   \centering
   \includegraphics[width=2.75in]{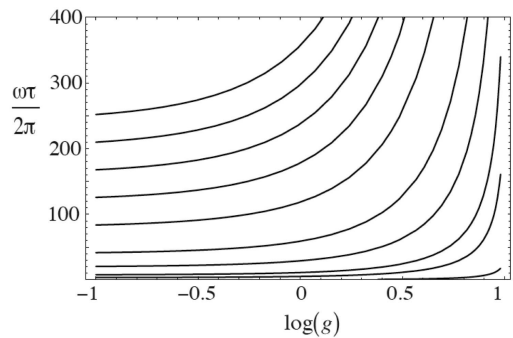} 
   \caption{Gate time under the assumption of adiabatic evolution as a function of $g$ for different values of $n$ in Eq. (\ref{adiabaticcondition}): $n= 1,5,10,25,50,100,200,250,300$.  Note, for $n$ small, the short gates times give very low fidelities due to diabatic effects in the true evolution (see Fig. 3).  The divergence of these curves as $g$ becomes large follows from the effect of the Tonks-Girardaeu regime.}
   \label{fig2}
\end{figure}

Though gates times can be calculated based on Eq. (\ref{adiabaticcondition}), there is no guarantee that these times are consistent with adiabatic evolution.  To evaluate the performance of this protocol, we must numerically solve the full time-dependent Schr\"{o}dinger equation.  Expanding in the adiabatic basis $\left\{\ket{\alpha(t)}\right\}$ and going to the rotating frame of the dynamical phase, the probability amplitudes evolve according to,
\begin{equation}
\dot{c}_{\alpha} = -\sum_{\beta} \bra{\alpha} \frac{\partial}{\partial t} \ket{\beta} e^{-i(\phi_{\beta}-\phi_{\alpha})}  c_{\beta} .
\end{equation}
In our numerical integration, we take a finite basis constructed from the span of two-particle orbitals with a maximum of two quanta of vibration in the separated wells.  The trajectory, $d(t)$, is chosen as a (suboptimal) linear ramp so that the $\sqrt{\text{SWAP}}$ is achieved according to the adiabatic dynamic phase evolution.  

We seek the fidelity of operation representing the overlap between final and target states.   Since quantum information is stored in the nuclear spin, we trace over the vibrational excitations.  For identical particles, one must take care so as not to artificially introduce entropy associated with (anti)symmetrization; subsystems must be defined by left/right localized atoms, not by particle label \cite{fermi_entanglement}. The $4 \times 4$ reduced density operator for the qubits is thus,
\begin{equation}
\rho^{\text{spin}}_{x'y',xy} (\tau) = \sum_{n,n'} \bra{vac}\hat{f}^{\phantom\dagger}_{L_n,x'}\hat{f}^{\phantom\dagger}_{R_{n'},y'} \hat{\rho}(\tau) \hat{f}^{\dagger}_{L_n,x}\hat{f}^{\dagger}_{R_{n'},y} \ket{vac},
\end{equation}
where $L_n$ and $R_n$ are left/right localized wavepackets with $n$ units of vibration, and $x$ and $y$ denote 0 or 1 for the logical basis.  In general, this density matrix will not have unit trace since there is finite probability of finding two atoms in the same well.  Double occupancy represents an infidelity in the operation.  The remaining infidelity arises from an error in reaching the target two-qubit spin state. We define this target as the two-atom spin state reached by adiabatic evolution for time $\tau$, where the initial state is the logical basis state $\ket{0,1}$.  The other possible initial logical basis states $\ket{0,0}$ and $\ket{1,1}$ evolve to good approximation by the unit operator and do not significantly contribute to infidelities when overall performance is good.  Our fidelity is the overlap between the chosen target and the two-atom spin state as calculated by the numerical solution under full evolution by the time-dependent Schr\"{o}dinger equation,
\begin{equation}
\mathcal{F} = \bra{\psi_{\text{target}}(\tau)} \hat{\rho}_{\text{spin}}(\tau) \ket{\psi_{\text{target}}(\tau)}.
\label{spinfidelity}
\end{equation}
When the time $\tau$ is chosen to satisfy Eq. (\ref{adiabaticcondition}), plotted in Fig 2, this is the fidelity for achieving the desired gate, $\ket{\psi_{\text{target}}}=\sqrt{\text{SWAP}}\ket{0,1} = (\ket{0,1}+i\ket{1,0})/\sqrt{2}$.   

Figure 3 shows the fidelity of adiabatic operation as a function of gate time and interaction strength.  Though high fidelity can be achieved for a wide range of parameters, robust operation will require high interactions and long time evolution.  As an example, for a coupling strength of $g=8$ and gate time of $\tau = 36 (2\pi/\omega)$, the fidelity to implement a $\sqrt{\text{SWAP}}$ is 0.982. Though not necessarily contributing to gate infidelity, vibrational heating can occur due to nonadiabatic dynamics.  We can calculate the mean vibration excitation in one atom after the operation according to
\begin{equation}
\left< n \right> = \sum_{n}  n\sum_{n',x,y} \bra{vac}\hat{f}^{\phantom\dagger}_{L_n,x}\hat{f}^{\phantom\dagger}_{R_{n'},y} \hat{\rho} \hat{f}^{\dagger}_{L_n,x}\hat{f}^{\dagger}_{R_{n'},y} \ket{vac} . 
\end{equation}
For the example above, the interaction will heat each atom to have  $\left< n \right> = 0.016$ quanta of vibration, on average.

\begin{figure}[htbp] 
   \centering
   \includegraphics[width=3.5in]{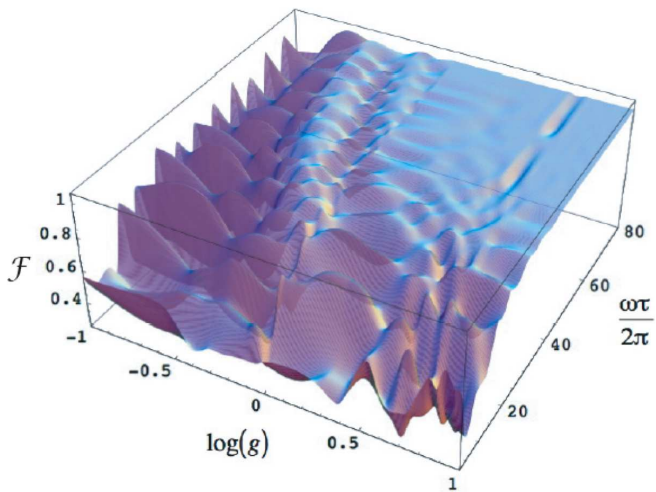} 
   \caption{Fidelity of adiabatic operation starting in the logical basis state $\ket{0,1}$, according to Eq. (\ref{spinfidelity}) as a function of $g$ and gate time $\tau$.  When $\tau$ and $g$ are chosen according to Eq. (\ref{adiabaticcondition}), this corresponds to the fidelity of the $\sqrt{\text{SWAP}}$ gate.}
   \label{fig3}
\end{figure}
The analysis above assumed initial preparation of the atoms in the vibrational ground state. We can study the tolerance to finite temperature in the initial state by averaging over a Boltzmann distribution of vibrational excitations.  The fidelity as a function of $\left< n(0) \right>$ falls off approximately linearly, $\mathcal{F}\left( \left< n(0) \right> \right) \sim -1.78 \left< n(0) \right> + \mathcal{F}_0$. The particular value of $\mathcal{F}_0$ depends on the parameters, as shown in Fig. 3.  A necessary condition for high fidelity is that initial vibrational excitation satistify $\left< n(0) \right> < 0.01$ or $kT/\hbar\omega<0.1$.  This extreme cooling requirement limits the practically of implementing this protocol in the lab.  Our focus here is on the proof-of-principle protocol, far from optimal, and generally very slow. Note, however, because the atom's nuclear spin is decoupled from electronic states, in principle damping forces can be introduced which maintain atoms near the ground vibrational state without decohering the qubit.  Such refrigeration can be introduced by laser cooling that does not flip nuclear spin \cite{Reichenbach} or through immersion in a superfluid where  sympathetic cooling can occur \cite{BECcooling}.  More optimal protocols will be studied in future work.

In conclusion, we have presented a new protocol for entangling qubits based on an exchange blockade between identical particles. Though we have studied this protocol in the context of identical spin-1/2 fermions, our result generalizes for an arbitrary spin, qubits or qudits, Bose or Fermi, and is not restricted to elements with electron and nuclear spin decoupled.  An equivalent evolution will arise if, in the absence of symmetrization for identical particles, the scattering phase shift is independent of logical basis state.  For the case of $^1S_0$ elements with an arbitrary nuclear spin, we can define $2s+1$ logical states \{$\ket{n,m}$\}  which will undergo a $\sqrt{\text{SWAP}}$ operation, $\ket{n,m}\Rightarrow\ket{n,m}+i\ket{m,n}$, as a consequence of a cold collision and the exchange blockade.  As another example, two ultracold  $^{87}$Rb atoms (bosons) will incur a s-wave collisional phase shift essentially independent of the magnetic sublevel.  If one encodes a qubit in two sublevels, $\ket{0}=\ket{F,m}, \ket{1}=\ket{F',m'}$, the logical basis states are equivalent to Eq. (\ref{logicalbasis}), with bose-symmetry substitution, $\ket{\Psi_+} \Leftrightarrow \ket{\Psi_-}$.  The s-wave collisions now occur for the symmetric combination of internal hyperfine levels, but not for the antisymmetric combination.  An equivalent protocol of bringing identical atoms into the same well and then separating them will yield a $\sqrt{\text{SWAP}}$ \cite{Porto2}.  

\begin{acknowledgments}
We thank Chad Hoyt, Leo Hollberg, Trey Porto, Marco Anderlini, and Carl Williams for helpful discussions. Partially supported by ARDA~DAAD19-01-1-0648 and ONR~N00014-03-1-0508.
\end{acknowledgments}

\end{document}